\documentclass[12pt,a4paper]{article}
\usepackage{amsmath}
\usepackage[dvips]{graphicx}
\input epsf
\usepackage{times}

\begin{document}
\begin{titlepage}
\title{Spin correlations due to antishadowing}
\author{S.M. Troshin\\[1ex]
\small  \it Institute for High Energy Physics,\\
\small  \it Protvino, Moscow Region, 142280, Russia} \normalsize
\date{}
\maketitle
\begin{abstract}
The effects of antishadowing related to the spin correlations of
particles in multiparticle production are discussed. It is shown
that significant spin correlations should be expected at the LHC
energies.

\end{abstract}
\end{titlepage}
\section*{Inroduction}
The importance of the spin degrees of freedom in multiparticle
production  has been recognized long time ago \cite{mich}. That
time it was understood that the popular models of particle
production were not able to reproduce (through unitarity) elastic
scattering data without account for the effects related to the
coherence of  spins of the final particles \cite{bial}. Later Chou
and Yang   have made conclusion on the correlations between spins
of final particles on the basis of their geometrical model of
particle production \cite{chyn}. However, those ideas have not
obtained appropriate attention since a dedicated experimental spin
studies with polarized beams and targets provide much more
information. During recent years a number of significant and
unexpected spin effects were discovered. They demonstrated that
the spin degrees of freedom will be important even at TeV energy
scale. With the start of the RHIC spin program \cite{rhic}, spin
studies moved again to the forefront of high energy physics.
However there will be no experimental possibilities at the LHC
similar to ones RHIC has, but the importance of spin effects will
definitely remain at such high energies and they will provide
important spin correlations. In this note we provide a ground for
the above statement, i.e. we show that the phenomena of
antishadowing \cite{uspl} expected at the LHC energies
\cite{petr,lasl} will be associated with strong coherent spin
effects: spins of the final particles should be aligned and this
alignment will have an important experimental signatures.

\section{Antishadowing and angular momentum conservation}
Multiparticle production in the model with antishadowing has been
studied in \cite{jpg}. It should be noted that antishadowing is a
result of the specific form of amplitude unitarization,
$U$--matrix or rational form of unitarity implementation at high
energies. It is characterized by the fact that beyond the
threshold value of energy  the inelastic overlap function
\[
\eta(s,b)\equiv\frac{1}{4\pi}\frac{d\sigma_{inel}}{db^2}
\]
has a peripheral impact parameter dependence. At the LHC energies
this peripheral profile with maximum at $b=R(s)$, where $R(s)$ is
the interaction radius, is predicted to be very prominent
\cite{petr}. The eikonal form of unitarization does not lead to
such peripheral picture. The difference between two unitarization
schemes can be clearly seen in Fig. 1.
\begin{center}
\begin{figure}[hbt]
\hspace*{2cm}\epsfxsize= 100  mm  \epsfbox{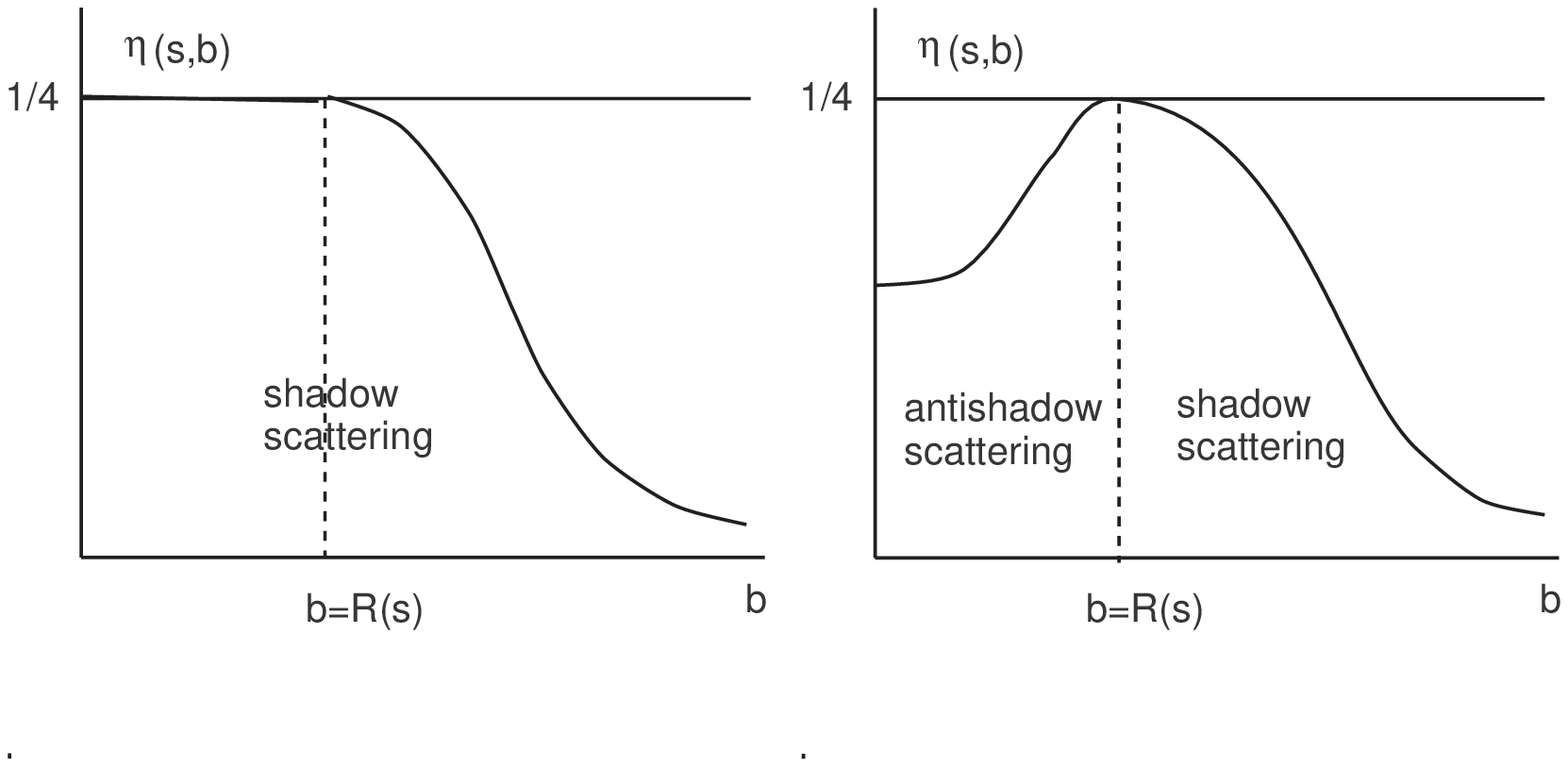}
\caption{Impact parameter dependence of the inelastic overlap
function
 in the eikonal unitarization scheme (left panel) and in the unitarization scheme
 with antishadowing (right panel).}
 \end{figure}
\end{center}
Thus, we can conclude that the region around $b=R(s)$ has the
highest probability of the multiparticle production in case of
antishadowing. Antishadowing leads to suppression of particle
production at small impact parameters and the main contribution to
the integral  multiplicity $\bar n(s)$ comes from the region of
$b\sim R(s)$ according to the relation:
\begin{equation}\label{mm}
\bar n(s)= \frac{\int_0^\infty  \bar n
(s,b)\eta(s,b)bdb}{\int_0^\infty \eta(s,b)bdb}.
\end{equation}
Thus due to peripheral form of the inelastic overlap function the
secondary particles will be mainly
 produced at impact parameters $b\sim R(s)$ and this will lead to imbalance
 between orbital angular momentum in the initial and final states since particles
 in the final state will carry  out large orbital
 angular momentum.
To compensate the imbalance in the orbital momentum
  spins of secondary particles should  become
  lined up, i.e. the spins of the produced particles should demonstrate
   significant
  correlations when the antishadow scattering mode appears.

  Now we will estimate the imbalance between the
  initial and final states. We  use for  the model
  for multiparticle production developed in \cite{jpg},
where  overlapping and interaction of peripheral clouds   occur at
the first stage of hadron interaction. As a result, massive
virtual quarks appear in the overlapping region. Massive virtual
quarks play a role of scatterers for the valence quarks in elastic
scattering; those quarks are transient ones in this process: they
are transformed back into the condensates of the final hadrons.
However they can be hadronized and this leads to production of
secondary particles in the central region.
   Moreover, valence quarks can
excite a part of the cloud of the virtual massive quarks and these
virtual massive
 quarks will also subsequently fragment into the multiparticle
final states. The latter mechanism is responsible for the
particle production in the fragmentation region and should lead to
strong correlations between secondary particles.

 Since we are dealing with constituent quarks, it is natural to expect
direct proportionality between  multiplicity of secondary
particles and number of virtual massive quarks appeared in the
intermediate state  (due to both mechanisms of multiparticle
production) in collision of the hadrons with given impact
parameter, i.e.:
\begin{equation}\label{mmult}
\bar n (s,b)=\alpha (n_{h_1}+n_{h_2})N_0(s)D_F(b)+ \beta
N_0(s)D_C(b),
\end{equation}
with  constant factors $\alpha$ and $\beta$, $n_{h_1}$ and
$n_{h_2}$ being the numbers of valence constituent  quarks in the
initial hadrons, $b$ is the impact parameter,
$N_0(s)\propto\sqrt{s}$, and the functions $D_F(b)$ and $D_C(b)$
are related to the distributions of matter in the valence
constituent quarks and of the condensate  in the initial hadrons
\cite{jpg}. We will consider symmetrical case of
$pp$--interactions. For this process the model leads to the equal
mean multiplicities of secondary particles in the forward and
backward hemispheres, i.e. $\bar n_F(s)=\bar n_B(s)$ and very
small longitudinal momentum transfer is expected between two
sides. In that sense this model is very similar to the model
developed by Chou and Yang \cite{chyn}, analysis of experimental
data performed in this paper confirms the assumption of small
longitudinal momentum exchange.

We can consider separately particles production in the forward and
backward hemispheres \cite{webber}. Let us consider for example
particles produced in the forward hemisphere. The  orbital angular
momentum in the initial state can be estimated as
\begin{equation}\label{li}
L_i\simeq \frac{\sqrt s}{2}\frac{R(s)}{2},
\end{equation}
where $R(s)$ is the interaction radius. The orbital angular
momentum in the final state is then
\begin{equation}\label{lfi}
L_f\simeq \frac{\bar n (s)\bar x_L(s)\sqrt s}{4}\frac{R(s)}{2},
\end{equation}
where we have taken into account that $\bar n (s)/2$ particles
with the average fraction of their longitudinal momentum
$\bar x_L(s)$ are produced at the impact parameter $R(s)/2$ due to
antishadowing (cf. Fig. 1). The average fraction of longitudinal momentum
$\bar x_L(s)$  according to the hypothesis
 of limiting fragmentation \cite{yen} would not decrease with energy. Thus
 we arrive to the negative imbalance of the orbital angular momentum
\begin{equation}\label{lf}
\Delta L=L_i-L_f\simeq \frac{\sqrt s}{2}\frac{R(s)}{2}(1- \frac{\bar n (s)\bar x_L(s)}{2}),
\end{equation}
i.e.
\begin{equation}\label{lfd}
\Delta L \simeq -\frac{\sqrt s}{2}\frac{R(s)}{2}\frac{\bar n (s)\bar x_L(s)}{2}.
\end{equation}
This negative $\Delta L$ should be compensated by
 the total positive spin $S$ of final particles (since the particles in the initial state
 are unpolarized)
\begin{equation}\label{sp}
S=-\Delta L
\end{equation}
This  spin alignment of produced particles appears in the
antishadowing mode when the particles are produced in the region
of impact parameters $b=R(s)$.

The vector of spin $\vec{S}$ lies in the transverse plane but it
cannot be detected through the transverse polarization of single
particle due to azimuthal symmetry of the production process
(integration over azimuthal angle $\varphi$). However, this effect
can be traced measuring transverse spin correlations of two
particles $\langle s_i s_j\rangle $.  The most evident way to
reveal this effect is to perform the measurements of the spin
correlations of hyperons whose polarizations can be extracted from
the angular distributions of their weak decay products.

Spin correlation should be stronger for the light particles and
weakening is expected for heavy particles since they should be
produced at smaller values of impact parameters.

\section*{Conclusion}
In conclusion we would like to emphasize the main point of this
note. We expect appearance at the LHC energies strong spin
correlations of final particles as a result of the prominent
antishadowing.

\section*{Acknowledgements}
The author is grateful to N.E. Tyurin for the helpful discussions
and comments and to V.A. Petrov for the interest in the result of
this work.

\end{document}